\documentclass[twocolumn,prb,showpacs]{revtex4}

\begin{document}
\topmargin-1.0cm

\title {Anisotropic splitting of intersubband spin plasmons in quantum wells with bulk and structural inversion asymmetry}
\author {C. A. Ullrich}
\affiliation {Department of Physics,  University of
Missouri-Rolla, Rolla, Missouri 65409}
\author {M. E. Flatt\'e}
\affiliation {Optical Science and Technology Center and Department of Physics and Astronomy, University of Iowa,
Iowa City, Iowa 52242}
\date{\today}

\begin{abstract}

In semiconductor heterostructures, bulk and structural inversion asymmetry and spin-orbit coupling induce a
{\bf k}-dependent spin splitting of valence and conduction subbands,
which can be viewed as being caused by momentum-dependent crystal magnetic fields.
This paper studies the influence of these effective magnetic fields on the intersubband 
spin dynamics in an asymmetric $n$-type GaAs/Al$_{0.3}$Ga$_{0.7}$As quantum well.
We calculate the dispersions of intersubband spin plasmons using linear response theory.
The so-called D'yakonov-Perel' decoherence mechanism is inactive for collective intersubband excitations,
i.e., crystal magnetic fields do not lead to decoherence of spin plasmons. Instead, 
we predict that the main signature of bulk and structural inversion asymmetry in intersubband
spin dynamics is a three-fold, anisotropic splitting of the spin plasmon dispersion.
The importance of many-body effects is pointed out, and conditions for experimental observation
with inelastic light scattering are discussed.
\end{abstract}

\maketitle

\newcommand{\scy}{\rm \scriptscriptstyle}
\newcommand{\qp}{q_{||}}
\newcommand{\rp}{r_{||}}
\newcommand{\kp}{k_{||}}
\newcommand{\qqp}{{\bf q}_{||}}
\newcommand{\rrp}{{\bf r}_{||}}
\newcommand{\kkp}{{\bf k}_{||}}
\newcommand{\bia}{^{\rm \scriptscriptstyle BIA}}
\newcommand{\sia}{^{\rm \scriptscriptstyle SIA}}
\newcommand{\ks}{^{\scriptscriptstyle \rm KS}}
\section{introduction} \label{sec:intro}

The field of spintronics \cite{awschalombook,wolf} is based on the concept of
exploiting the spin degree of freedom of carriers to develop new features and functionalities 
for solid-state devices. Many of the proposed new applications rely, in one form or another, 
on manipulating nonequilibrium spin coherence. Two characteristic times, $T_1$ and $T_2$, 
provide a quantitative measure for the magnitude and persistence of spin coherence: $T_1$ 
describes the return to equilibrium of a non-equilibrium spin population, and $T_2$ measures 
the decay of transverse spin order.\cite{orientation,flatte}

In this paper, we consider electronic spin dynamics in quantum wells 
not {\em within one} but {\em between two} subbands (we will limit the discussion here 
to conduction subbands). 
One motivation for this work is that intersubband (ISB) {\em charge} dynamics in quantum
wells is currently of great experimental and theoretical interest, \cite{book1,book2}
since electronic ISB transitions are the basis of a variety of new devices operating in the terahertz
frequency regime, such as detectors, \cite{detector} modulators \cite{modulator}
and quantum cascade lasers. \cite{laser,kohler}
In view of this, it seems worthwhile to explore ISB {\em spin} dynamics as a possible
route towards novel applications in the terahertz regime. 

Characteristic times for ISB dynamics are also referred to as $T_1$ and $T_2$, whether or
not decay of spin coherence is involved.\cite{helm} Population decay
from an excited to a lower conduction subband is characterized by an ISB relaxation time 
$T_1^{\scy ISB}$, and loss of coherence of collective ISB excitations is
measured by a decoherence time $T_2^{\scy ISB}$.
Much has been learned recently about ISB charge-density excitations (CDE's, also called charge plasmons).
The two characteristic times were measured experimentally for CDE's in quantum wells \cite{heyman,williams} 
and were found to differ substantially at low
temperatures, $T_2^{\scy ISB}$ being three orders of magnitude smaller than $T_1^{\scy ISB}$.
The reason is that CDE relaxation characterized by $T_1^{\scy ISB}$ proceeds mainly via phonon emission 
and is thus slowed down by an energy bottleneck for acoustic phonons with small momenta and all optical
phonons. 

In contrast to $T_1^{\scy ISB}$, decoherence ($T_2^{\scy ISB}$) of collective ISB CDE's in quantum wells
is determined by a complex interplay of a variety of different scattering
mechanisms, whose relative importance is not {\em a priori} obvious. 
In recent experimental \cite{jon} and theoretical \cite{PRL,TCDFT,unuma} work, it was found
that the linewidth of (homogeneously broadened) ISB charge plasmons in a wide 
GaAs/Al$_{0.3}$Ga$_{0.7}$As quantum well (where phonon scattering is not important)
is determined mainly by interface roughness and electronic many-body effects. 

The question now arises which physical mechanisms govern the decoherence of ISB
{\em spin-density} excitations (SDE's, or spin plasmons). Near room temperature the {\em intra}subband
electron spin decoherence in semiconductors is to a large extent determined by the 
D'yakonov-Perel' mecha\-nism, \cite{DP1,DP2,DK} which involves spin precession of
carriers with finite crystal momentum $\bf k$ in an effective $\bf k$-depen\-dent crystal magnetic
field in inversion-asymmetric materials \cite{dresselhaus,rashba} and implies $T_1$ and $T_2$ are comparable.
Using this theory, Lau {\em et al.}\cite{lau,lau1} achieved good agreement with experimental
spin decoherence times, without including electron-electron interactions (see also Refs. 
\onlinecite{averkiev,bournel,karimov}). Thus, it is clear that crystal magnetic fields are a fundamentally 
important factor in determining $T_1$ and $T_2$ in intrasubband spin dynamics. 
However, the role of the crystal magnetic fields for $T_1^{\scy ISB}$ and $T_2^{\scy ISB}$ has so
far remained an open question.

This paper addresses the influence of the $\bf k$-dependent crystal magnetic field in 
semiconductor quantum wells on ISB spin plasmons. Building on a formal framework introduced 
previously,\cite{ullrichflatte}  we use (time-dependent) density functional theory to describe static 
and dynamic many-body effects. We extend our previous result\cite{ullrichflatte} that due to their 
collective nature the ISB spin plasmons are robust against D'yakonov-Perel' decoherence. 
The main reason for this suppression is that the {\bf k}-average of the crystal magnetic field
vanishes for an electron occupation function symmetric in {\bf k}, due to time-reversal invariance.
This result was previously reported only for Rashba effective magnetic fields, and we now include
effective magnetic fields arising from both bulk inversion asymmetry and Rashba effects.

The general structure of the ISB plasmon dispersions consists of a CDE branch above 
and three SDE branches below the region of single-particle excitations.\cite{ryan, marmorkos}
In the absence of any magnetic fields, the three SDE branches (one longitudinal, two transverse)
are degenerate. As we will show, the crystal magnetic field further splits the three SDE dispersion
relations in an anisotropic way.
We will predict that this effect should be observable with inelastic light scattering techniques.
\cite{scattering,abstreiter,pinczuk,gammon,gammon1} We mention that similar predictions were made, based on
Fermi liquid theory, by 
Mal'shukov {\em et al.},\cite{malshukov,brataas} who discussed ISB spin dynamics for model III-V quantum wells 
without including the Rashba effect. In the present paper, we will consider a realistic, asymmetrically
doped GaAs/$\rm Al_{0.3}Ga_{0.7}As$ quantum well system \cite{jusserand,wissinger} that features an
interesting interplay between bulk and structural inversion asymmetry.\cite{dresselhaus,rashba}

This paper is organized as follows. In Section \ref{secII} we
calculate the electronic ground state in modulation-doped 
quantum well conduction subbands, including spin-orbit coupling and many-body effects.
In Section \ref{secIII} we use a linear response formalism for the 
spin-density matrix, based on time-dependent density functional theory, to calculate the ISB
plasmon dispersions. We also discuss possibilities for experimental observation of the spin plasmons
with inelastic light scattering. Section \ref{secIV} contains our conclusions.

\section{Conduction subband spin splitting in quantum wells}\label{secII}

The first step is to calculate the momentum-dependent spin splitting of conduction subbands in a quantum 
well in the presence of bulk inversion asymmetry \cite{dresselhaus} (BIA) and structural inversion asymmetry (SIA)
or Rashba effect. \cite{rashba} A detailed account of the method is given in Ref. \onlinecite{ullrichflatte}.
We use a $2\times 2$ conduction subband Hamiltonian, which can be obtained by reduction from a multi-band
$\bf k \cdot p$ Hamiltonian in a standard way. 
\cite{rossler,warburton,eppenga,andrada1,andrada2,pfeffer1,pfeffer1a,pfeffer2,pfeffer2a}
This results in the following two-component form of the electron wavefunction for the $j$th subband:
\begin{equation} \label{II.1}
\Psi_{j{\bf q}_{||}}({\bf r}) = e^{i \qqp \rrp} \left( \begin{array}{c}
\psi_{j\uparrow}(\qqp,z) \\ \psi_{j\downarrow}(\qqp,z) \end{array} \right),
\end{equation}
where $\rrp$ and $\qqp$ are in-plane position and wave vectors, and $z$ is the direction of
growth of the quantum well.
The envelope functions $\psi_{j\sigma}$ are obtained from a two-component
effective-mass Kohn-Sham equation, which in its most general form reads as follows: 
\begin{eqnarray}  \label{II.2}
\lefteqn{ \hspace{-2.5cm}
\sum_{\beta = \uparrow,\downarrow} \left( 
\hat{h}\:\delta_{\alpha \beta} 
+ v^{\rm ext}_{\alpha \beta}(z) 
+ \hat{H}^{\rm so}_{\alpha \beta}(z) 
+ v^{\rm xc}_{\alpha \beta}(z) \right) \psi_{j\beta} (\qqp,z)} \nonumber\\
&=& E_{j\qqp} \,\psi_{j\alpha} (\qqp,z)\;, 
\end{eqnarray}
where $\alpha = \uparrow,\downarrow$, and
\begin{equation}\label{II.3}
\hat{h} = - \, \frac{d}{ dz} \, \frac{\hbar^2}{2 m^*(z)} \, \frac{d}{dz}
+ \frac{\hbar^2 q_{||}^2}{2 m^*(z)} +v_{\rm conf}(z) + v_{\rm H}(z) .
\end{equation}
We use here the simplifying assumption of energy-independent effective masses $m^*$,
which is sufficient for the GaAs/AlGaAs system studied in this paper.
$v_{\rm conf}(z)$ is the confining bare quantum well potential (e.g., a square well).
The Hartree potential $v_{\rm H}(z)$  describes the classical Coulomb potential due to the electron ground-state
density $n(z)$ and to the density of positive donor impurities $n_i(z)$:
\begin{equation} \label{II.4}
\frac{d^2 v_{\rm H}(z)}{dz^2} = - 4\pi {e^*}^2 [n(z) - n_i(z)] \;, 
\end{equation}
where $e^* = e/\sqrt{\epsilon}$ is the effective charge (the static
dielectric constant is taken as $\epsilon = 13$ throughout the system).

In this paper, we consider the case without any externally applied static electric or magnetic
fields, so that $v^{\rm ext}_{\alpha \beta}(z) = 0$. As a consequence, the exchange-correlation (xc)
potential becomes diagonal in the spins: $v^{\rm xc}_{\alpha \beta}(z) = v_{\rm xc}(z) \delta_{\alpha \beta}$.
We take a standard local-density approximation (LDA) for $v_{\rm xc}(z)$.\cite{ullrichflatte}

Intrinsic conduction band spin splitting is caused by  spin-orbit interaction, but may originate from various
sources.\cite{flatte} In the following, we will consider BIA and SIA
contributions:  $\hat{H}^{\rm so}_{\alpha \beta}= \hat{H}\bia_{\alpha \beta}+\hat{H}\sia_{\alpha \beta}$.
The BIA term depends on the direction of growth of the quantum well, \cite{DK,eppenga}
which we here take along [001].

The spin-orbit Hamiltonian can be written as
\begin{equation} \label{II.5}
\hat{H}^{\rm so} = \vec{\sigma} \cdot {\bf B}_{\rm eff}  \:,
\end{equation}
where $\vec{\sigma}$ is the vector of the Pauli spin matrices, and ${\bf B}_{\rm eff}$ acts as an in-plane effective 
magnetic field. Some simplification is achieved using a
perturbative treatment,\cite{eppenga} and one arrives at
\begin{equation} \label{II.6}
{\bf B}_{\rm eff} = \left( \begin{array}{c}
\langle \gamma \rangle q_x q_y^2  - \langle \gamma \hat{q}_z^2 \rangle q_x  + \langle \alpha \rangle q_y \\
-\langle \gamma \rangle q_y q_x^2 + \langle \gamma \hat{q}_z^2 \rangle q_y  - \langle \alpha \rangle q_x \\
0
\end{array} \right).
\end{equation}
This leads to 
\begin{equation} \label{II.7}
\hat{H}^{\rm so}_{\uparrow\uparrow} = \hat{H}^{\rm so}_{\downarrow\downarrow} = 0 \;, \qquad
\hat{H}^{\rm so}_{\uparrow\downarrow} = (\hat{H}^{\rm so}_{\downarrow\uparrow})^\dagger = \hbar \Omega\;, 
\end{equation}
where
\begin{equation} \label{II.8}
\hbar\Omega = \langle \gamma \rangle q_x q_y (q_y + i q_x) 
-\langle \gamma \hat{q}_z^2 \rangle (q_x + i q_y) 
+  \langle\alpha \rangle(q_y + i q_x) \:.
\end{equation}
$\gamma$ and $\alpha$ are material parameters determining the BIA and SIA effects, and are
position dependent in a quantum well. Thus, we have
\begin{equation}
\langle \gamma \rangle = \int dz\: \gamma(z) \, |\psi_j(z)|^2 \;,
\end{equation}
the analogous definition for $\langle \alpha \rangle$, and 
\begin{equation}
\langle \gamma \hat{q}_z^2 \rangle = - \int dz \: \psi_j(z)\: \frac{d}{dz} \: \gamma(z) \frac{d}{dz} \:
\psi_j(z) \;.
\end{equation}
(notice the dependence on the subband index $j$). Several authors
\cite{andrada1,andrada2,pfeffer1,pfeffer1a,pfeffer2,pfeffer2a} have pointed out the importance of a proper treatment of the 
discontinuities at the left and right quantum well interfaces, $z = z_{L,R}$.  Thus, we take
\begin{eqnarray}
-\langle \gamma \hat{q}_z^2 \rangle &=& \int ' dz \: \psi_j(z)\: \frac{d}{dz} \: \gamma(z) \frac{d}{dz} \:
\psi_j(z) \nonumber\\
&& {}+ \Gamma_L \psi_j(z_L) + \Gamma_R \psi_j(z_R),
\end{eqnarray}
where the prime on the integral means that infinitesimal regions around the interfaces are excluded in the 
integration, and the jumps $\Gamma_{L,R}$ are defined as
\begin{equation}
\Gamma_{L,R} = \left[\gamma(z) \frac{d\psi_j(z)}{dz}\right]_{z_{L,R}^+} - 
\left[\gamma(z) \frac{d\psi_j(z)}{dz}\right]_{z_{L,R}^-} .
\end{equation}
A similar treatment is carried out for the jumps in $\langle \alpha \rangle$, where
the Rashba coefficient $\alpha$ is given as \cite{andrada1,andrada2}
\begin{equation}
\alpha = \frac{d}{dz} \: \frac{\hbar^2}{2 m^*} \left[ \frac{\Delta}{3 E_g + 2 \Delta} - 
\: \frac{(\varepsilon_j - V)(2 E_g \Delta + \Delta^2)}{E_g(E_g+\Delta)(3 E_g+2 \Delta)} \right] .
\end{equation}
Here, $V = v_{\rm conf} + v_{\rm H} + v_{\rm xc}$ is the total potential, and
the $j$th subband energy $\varepsilon_j$ is defined as the energy above the minimum of $V$ inside the quantum well
($z_L < z < z_R$). We take a conduction band offset of 257.6 meV between GaAs and $\rm Al_{0.3}Ga_{0.7}As$.
All other material parameters determining $\gamma$ and $\alpha$ are summarized in Table  \ref{table1}.

\begin{table}
\caption{\label{table1} Material parameters for GaAs and $\rm Al_{0.3}Ga_{0.7}As$.}
\begin{ruledtabular}
\begin{tabular}{ccccc}
 & $E_g$ (eV) & $\Delta$ (eV) & $\gamma$ (eV \AA$^3$) & $m^*$  \\ \hline
GaAs &  1.519 & 0.34         & 24.12   &0.067  \\
$\rm Al_{0.3}Ga_{0.7}As$ & 1.921 & 0.322 & 18.03 & 0.092
\end{tabular}
\end{ruledtabular}
\end{table}

The two-component Kohn-Sham equation (\ref{II.2}) thus becomes
\begin{equation} \label{4.2}
\left( \begin{array}{cc} \hat{h} + v_{\rm xc} & \hbar\Omega \\
\hbar\Omega^* & \hat{h} + v_{\rm xc} \end{array}
\right)
\left( \begin{array}{c} \psi_{i\uparrow} \\ \psi_{i\downarrow} \end{array} \right) = 
E_{i \qqp} 
\left( \begin{array}{c} \psi_{i\uparrow} \\ \psi_{i\downarrow} \end{array} \right) ,
\end{equation}
where $ i = 1,2,3,\ldots$, and $\hat{h}$ and $\Omega$ are given by Eqs. (\ref{II.3}) and (\ref{II.8}).
Equation (\ref{4.2}) is solved by the following ansatz:
\begin{eqnarray}\label{4.4}
\psi_{sj\uparrow}(\qqp,z) &=& \frac{1}{\sqrt{2}} \: \varphi_j(z) \\
\psi_{sj\downarrow}(\qqp,z) &=& \frac{s}{\sqrt{2}} \: \frac{\Omega^*}{|\Omega|} \: \varphi_j(z)\;,
\label{4.5}
\end{eqnarray}
where we replaced the subband index $i$ by the pair of indices $\{sj\}$, such that
$s = (-1)^i$ and $j = (i+1)/2$ for $i$ odd and $j=i/2$ for $i$ even. In the absence of
the off-diagonal terms in Eq. (\ref{4.2}), i.e. for inversion symmetry and hence spin degeneracy
at each $\qqp$, $j$ simply labels the spin-degenerate pairs, and $s$ labels the eigenfunctions within each pair.

The $\varphi_j(z)$ are the solutions of a spin-inde\-pen\-dent effective-mass Kohn-Sham
equation:
\begin{equation}
\left[- \frac{d}{dz}\frac{\hbar^2}{2m^*(z)}\frac{d}{dz}+ v_{\rm conf} + v_{\rm H}+ v_{\rm xc} \right] 
\varphi_j = \varepsilon_j \varphi_j \;,
\end{equation}
where $\varepsilon_j$ are the energy levels of the associated, doubly degenerate, parabolic subbands.
The presence of the off-diagonal BIA and SIA terms in Eq. (\ref{4.2}), however,  lifts the 
spin degeneracy for $\qqp \ne 0$. Ignoring the slight non-parabolicity that comes from the
$z$-dependence of $m^*$, we obtain
\begin{equation}
E_{sj \qqp} = \varepsilon_j + \frac{\qp^2}{2m^*} + s \hbar |\Omega| \;, \qquad s=\pm 1\;,
\end{equation}
for the energy eigenvalues associated with the solutions (\ref{4.4}),(\ref{4.5}) of Eq. (\ref{4.2}),
where $m^*$ is the GaAs effective mass.

\begin{figure}
\unitlength1cm
\begin{picture}(5.0,5.5)
\put(-5.7,-9.8){\makebox(5.0,5.5){
\includegraphics{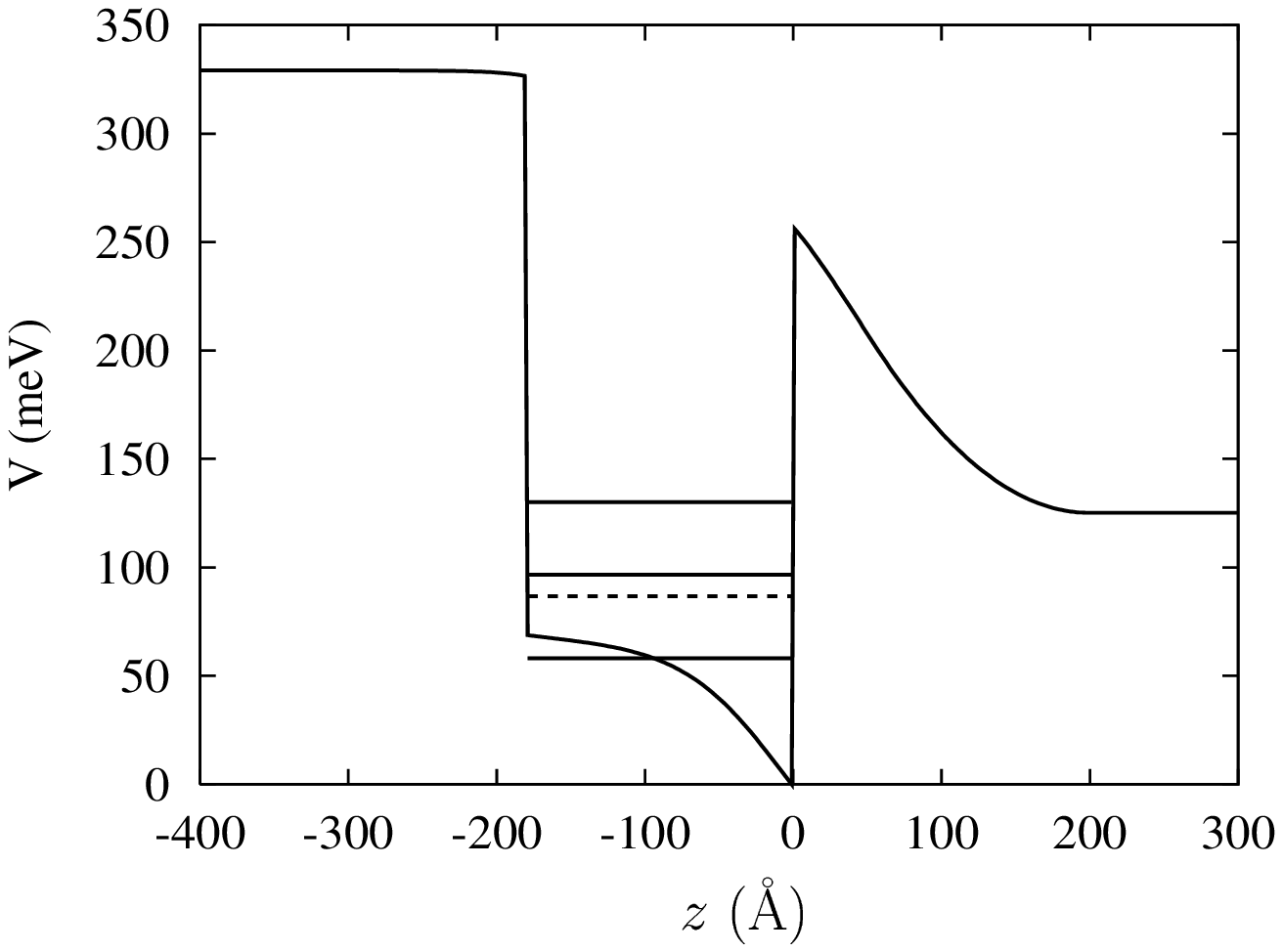}}}
\end{picture}
\caption{\label{figure1} Total potential $V$ and conduction subband levels $\varepsilon_j$
of an asymmetrically doped 180{\AA} GaAs/$\rm Al_{0.3}Ga_{0.7}As$ quantum well, with
electron density $N_s = 8 \times 10^{11} \: \rm cm^{-2}$. The dashed line indicates the
conduction band Fermi level.
}
\end{figure}

Fig. \ref{figure1} shows the conduction band edge and subband levels of an asymmetrically doped,
180 {\AA} wide, [001]-grown GaAs/$\rm Al_{0.3}Ga_{0.7}As$ quantum well.
The carrier sheet density is taken as $N_s = 8 \times 10^{11} \: \rm cm^{-2}$, 
and the same number of positive donor impurities is evenly distributed between
$50 \mbox{\AA} < z <  200 \mbox{\AA}$, which defines $n_i(z)$ in Eq. (\ref{II.4}).
It is seen from the figure that only the lowest subband is occupied.

\begin{figure}
\unitlength1cm
\begin{picture}(5.0,5.3)
\put(-5.7,-10.8){\makebox(5.0,5.3){
\includegraphics{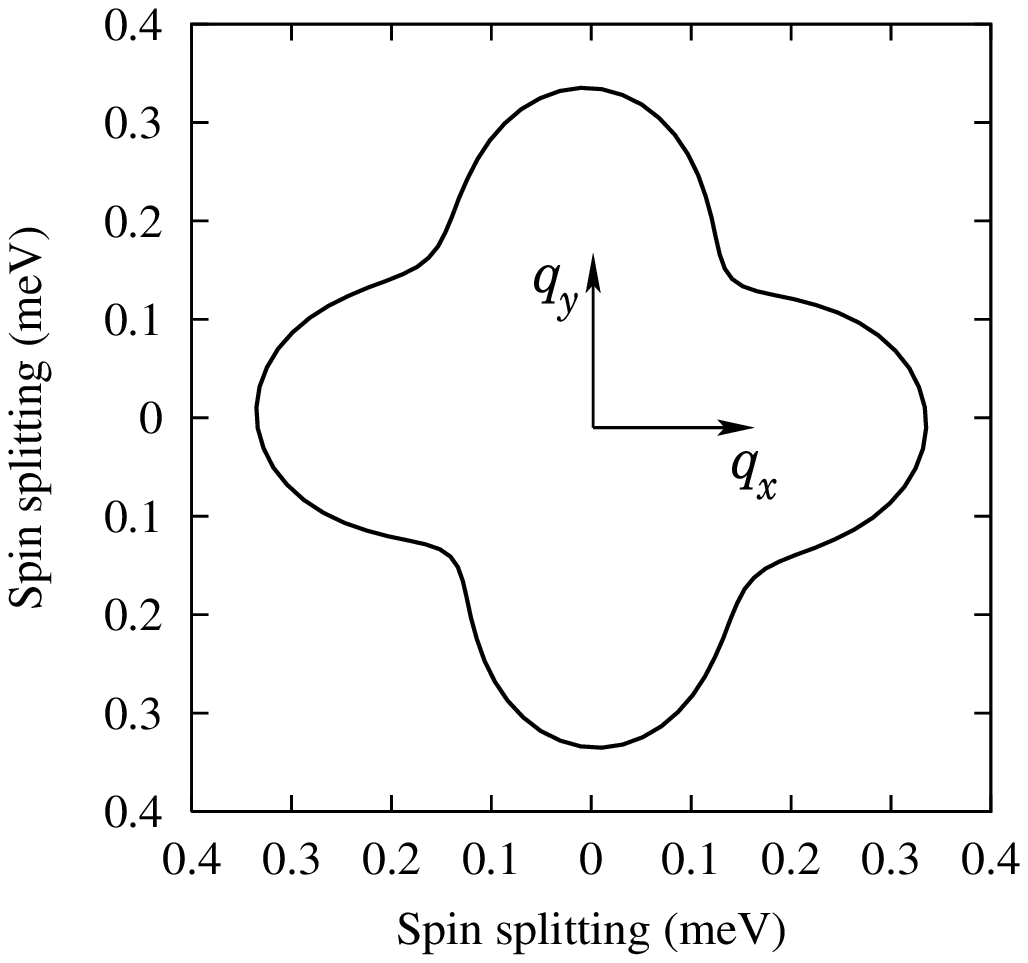}}}
\end{picture}
\caption{\label{figure2} Spin splitting at the Fermi level of the lowest conduction subband
of the quantum well from Fig. \ref{figure1}.
}
\end{figure}

The conduction subband splitting in the same quantum well system was studied before
by Jusserand {\em et al.} \cite{jusserand} using intrasubband Raman scattering, 
and theoretically by Wissinger {\em et al.} \cite{wissinger} using more sophisticated $8 \times 8$
and $14 \times 14$ band models. Our results, obtained with the simpler $2\times 2$ model, agree 
very well with these previous works. \cite{jusserand,wissinger} We illustrate in Fig. \ref{figure2}
the splitting of the lowest conduction subband at the Fermi level, where $q_F = 0.02 \mbox{\AA}^{-1}$.
There is a pronounced anisotropy: the spin splitting is 0.34 meV along [100], 0.23 meV along [110],
and 0.20 meV along $[1 \bar{1} 0]$.

\begin{figure}
\unitlength1cm
\begin{picture}(5.0,9.5)
\put(-5.5,-9.){\makebox(5.0,9.5){
\includegraphics{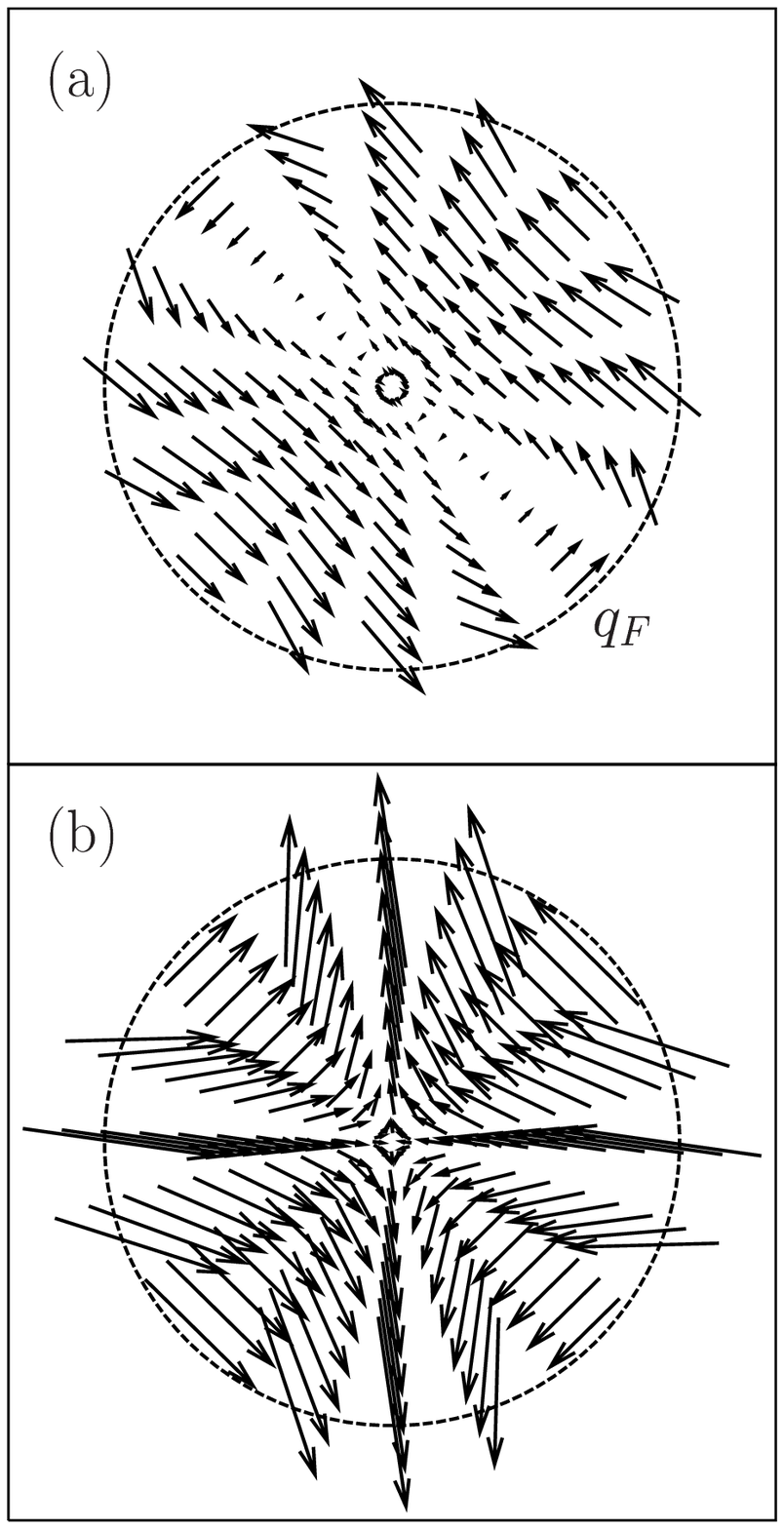}}}
\end{picture}
\caption{\label{figure3} Momentum-dependent crystal magnetic field ${\bf B}_{\rm eff}$ [Eq. (\ref{II.6})] for the first
(a) and second (b) subband of the quantum well of Fig. \ref{figure1}, for in-plane momenta $\qp < q_F$.
}
\end{figure}

Fig. \ref{figure3} shows the crystal magnetic fields ${\bf B}_{\rm eff}$ [Eq. (\ref{II.6})]
for the first and second subband of our quantum well. It can be seen how magnitude as well as 
direction of ${\bf B}_{\rm eff}$ strongly depend on the in-plane momentum $\qqp$, which illustrates the physical origin of
the D'yakonov-Perel' decoherence mechanism:\cite{DK} carriers with different momentum, initially in phase, 
precess at different rates, thus losing coherence. For our system, it appears as if the
crystal magnetic field in the lowest subband is dominated by the Rashba effect, whereas the
second (lowest unoccupied) subband is dominated by the Dresselhaus bulk term.

\section{Intersubband spin dynamics}\label{secIII}

\subsection{Formal framework}

Having calculated the subband energy levels and envelope functions, we now
consider ISB excitations.
Formal basis is the linear response theory for the
spin-density matrix that was developed earlier. \cite{ullrichflatte}
ISB charge and spin plasmons emerge as solutions of the response equation 
corresponding to collective CDE's and SDE's.

We denote the first-order change of the spin-density matrix by $n^{(1)}_{\sigma \sigma'}(\kkp,z,\omega)$,
and define $m_j^{(1)} = {\rm Tr} \left[\sigma_j \: \underline{n}^{(1)} \right]$,
$j=0,1,2,3$, where $\sigma_0$ is the $2\times 2$ unit matrix, and 
$\sigma_1,\sigma_2,\sigma_3$ are the Pauli matrices.
$m_0^{(1)}=n^{(1)}_{\uparrow \uparrow} + n^{(1)}_{\downarrow \downarrow} $ 
describes a collective CDE,  and
$m_3^{(1)}=n^{(1)}_{\uparrow \uparrow} - n^{(1)}_{\downarrow \downarrow}$ is a 
longitudinal SDE with respect to the $z$-axis. 
In terms of this choice of global spin quantization, 
$m_1^{(1)}=n^{(1)}_{\uparrow \downarrow} + n^{(1)}_{\downarrow \uparrow} $ 
and $m_2^{(1)}=i[n^{(1)}_{\uparrow \downarrow} - n^{(1)}_{\downarrow \uparrow}]$ appear as transverse SDE's
(or spin-flip excitations).

The spin-density-matrix response $n^{(1)}_{\sigma \sigma'}$ couples to spin-dependent potentials
$v^{(1)}_{\sigma \sigma'}$, which we can again combine as 
$V_j^{(1)} = {\rm Tr} \left[\sigma_j \: \underline{v}^{(1)} \right]$.
Thus, the CDE couples to an oscillating electric 
field polarized along the $z$-direction, associated with $V_0^{(1)}$.
The longitudinal SDE is excited by an oscillating magnetic field along $z$ 
associated with $V_3^{(1)}$, and the transverse SDE's are excited by oscillating magnetic fields 
along $x$ and $y$, associated with $V_1^{(1)}$ and $V_2^{(1)}$, respectively. 
We will discuss the corresponding selection rules for inelastic light scattering below.

In terms of these quantities, the linear response equation takes on the following $4\times 4$ matrix form:
\begin{equation}\label{III.1}
m_j^{(1)}(\kkp,z,\omega) = \sum_{k=0}^3 \int \! dz' \: \Pi^{\rm \scriptscriptstyle KS}_{jk}
(\kkp,z,z',\omega)  V_k^{(1)}(\kkp,z',\omega) .
\end{equation}
$\Pi\ks_{jk}$ is the non-interacting (Kohn-Sham) charge-spin response function, which was given in Appendix B
of Ref. \onlinecite{ullrichflatte}.
The $V_k^{(1)}$, in turn, are  sums of external perturbations and linearized
Hartree and xc terms:
\begin{eqnarray} \label{III.2}
V_k^{(1)}(\kkp,z,\omega) &=& V^{\rm ext}_k(\kkp,z,\omega)
\nonumber\\
&+&
\sum_{l=0}^3 
\int\!dz' \left[ \frac{2 \pi {e^*}^2}{\kp}\: e^{-\kp|z-z'|} \:\delta_{k0}\delta_{l0} \right.
\nonumber\\
&+&
 f^{\rm xc}_{kl}(\kkp,z,z',\omega)\bigg] m_l^{(1)}(\kkp,z',\omega) .
\end{eqnarray}
The xc kernels $f^{\rm xc}_{kl}$ are given in Appendix A of Ref. \onlinecite{ullrichflatte}.
Eq. (\ref{III.1}) can be rewritten formally exactly as
\begin{equation}\label{III.3}
m_j^{(1)}(\kkp,z,\omega) = \sum_{k=0}^3 \int \! dz' \: \Pi_{jk}
(\kkp,z,z',\omega)  V_k^{\rm ext}(\kkp,z',\omega) ,
\end{equation}
which defines the full charge-spin response function:
\begin{equation} \label{III.4}
{\bf \Pi} = \frac{{\bf \Pi}\ks}{ {\bf 1} - {\bf \Pi}\ks {\bf F}^{\rm Hxc}} \;.
\end{equation}
The elements of the matrix  ${\bf F}^{\rm Hxc}$ are given by
\begin{eqnarray} \label{III.5}
F^{\rm Hxc}_{kl}(\kkp,z,z',\omega) &=&  \frac{2 \pi {e^*}^2}{\kp}\: e^{-\kp|z-z'|} \:\delta_{k0}\delta_{l0} \nonumber \\
&+& f^{\rm xc}_{kl}(\kkp,z,z',\omega) \;.
\end{eqnarray}
The poles of ${\bf \Pi}\ks$ yield the single-particle excitation spectrum, whereas the poles
of $\bf \Pi$ are the charge and spin plasmons.

Experimental observation of ISB plasmons can be achieved with various methods of optical 
spectroscopy.\cite{vasko} Consider the following expression:
\begin{equation} \label{III.6}
\bar{\Pi}_{jk} (\kkp,\omega) = \int\! dz \! \int\! dz' \: z z' \: \Pi_{jk}(\kkp,z,z',\omega) \;,
\end{equation}
which entails the proper ISB dipole selection rules for interaction of the quantum well carriers
with electromagnetic waves. CDE's can be observed using photoabsorption
spectroscopy \cite{helm,heyman,williams}  or inelastic light scattering
\cite{scattering,abstreiter,pinczuk,gammon,gammon1}
in the so-called {\em polarized} geometry,  with cross sections given in both cases by
\begin{equation} \label{III.7}
\sigma_c(\kkp,\omega) \sim {\rm Im}\: \bar{\Pi}_{00}(\kkp,\omega) \;.
\end{equation}
SDE's, on the other hand, are seen with inelastic light scattering in the {\em depolarized} geometry, with cross sections
\begin{equation}\label{III.8}
\sigma_s(\kkp,\omega) \sim {\rm Im} \sum_{i,j=1}^3 P_i \bar{\Pi}_{ij}(\kkp,\omega) P_j^* \;.
\end{equation}
This expression implies orthogonal polarization vectors of the incident and scattered light,
$\vec{e}_{i,s}$, through $\vec{P} = \vec{e}_i \times {\vec{e}_s}^*$.
We also note that no definite selection rules exist for incoherent single-particle excitations,  \cite{burstein}
which thus show up in both $\sigma_c$ and $\sigma_s$. 

Finite linewidths of collective CDE's and SDE's in weakly disordered systems 
enter through the structure of $\Pi_{ij}$: either in the form of phenomenological scattering times
(which can essentially be put in by hand), or through microscopic approaches such as
the memory-function formalism. \cite{PRL} In the following, however, we will not be concerned with
dissipation mechanisms such as impurity, interface roughness, phonon or electron-electron scattering.

\subsection{Results and Discussion}

\begin{figure}
\unitlength1cm
\begin{picture}(5.0,5.5)
\put(-5.9,-9.2){\makebox(5.0,5.5){
\includegraphics{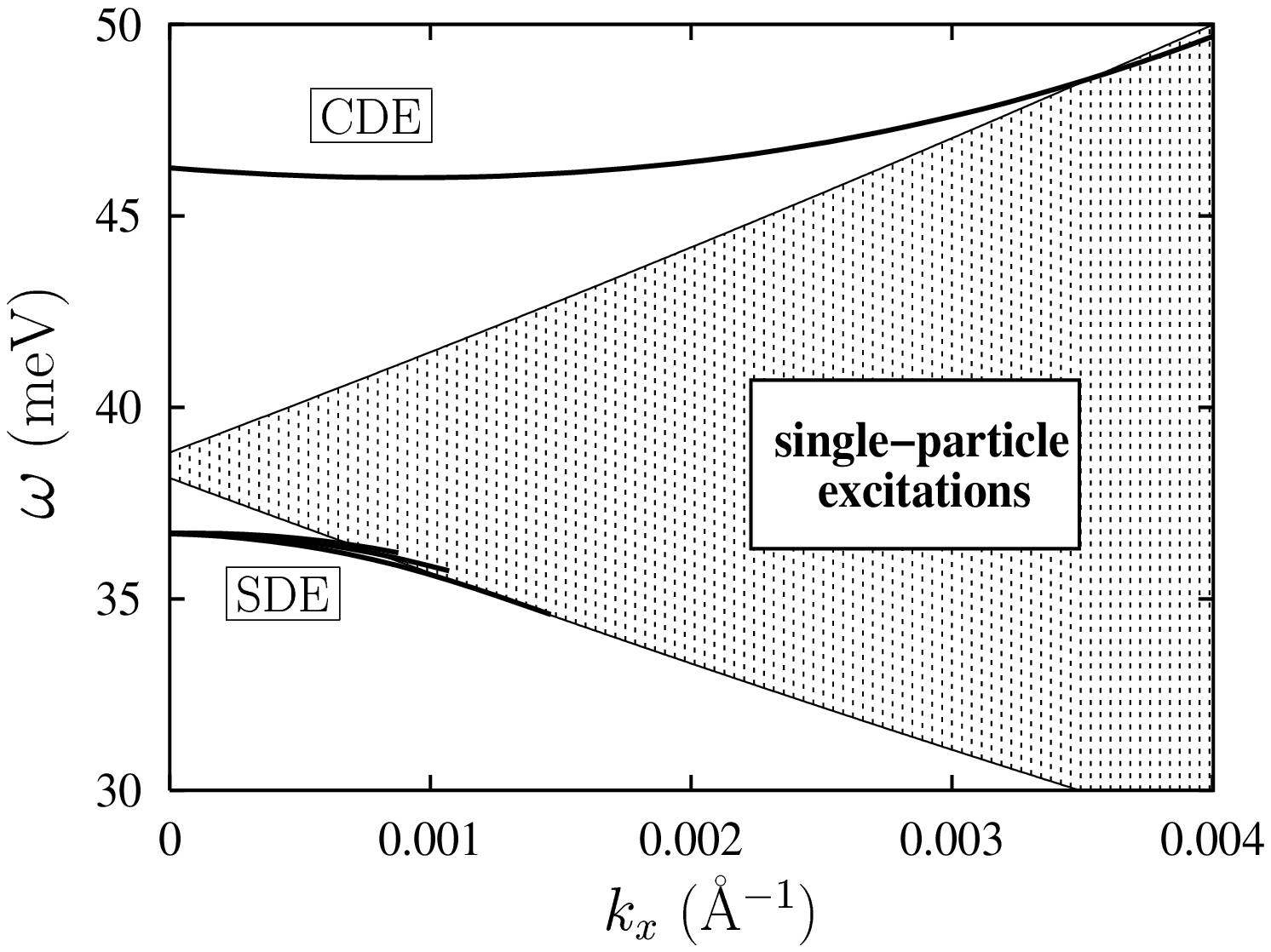}}}
\end{picture}
\caption{\label{figure4} ISB charge and spin plasmon wave vector dispersions in the
quantum well of Fig. \ref{figure1}, for ${\bf k} || [100]$.
}
\end{figure}

Fig. {\ref{figure4} shows the wave vector dispersions of the CDE and the SDE in our GaAs/$\rm Al_{0.3}Ga_{0.7}As$
quantum well. The direction of the plasmon wave vector is taken along [100].
The overall qualitative picture of the ISB plasmon dispersions is well known. \cite{ryan,marmorkos}
The shaded region indicates the particle-hole continuum: if the collective modes enter this
region, they rapidly decay into incoherent single-particle excitations (Landau damping).
The CDE lies above the single-particle region, and the SDE's lie below.
We mention that in inelastic light scattering experiments, ISB plasmon dispersions can be
measured up to a maximum wave vector transfer of $\kp \sim 0.002 \: \mbox{\AA}^{-1}$,
by rotating the sample with respect to the incident and detected wavevector of the light. \cite{gammon1}

\begin{figure}
\unitlength1cm
\begin{picture}(5.0,5.5)
\put(-5.9,-9.1){\makebox(5.0,5.5){
\includegraphics{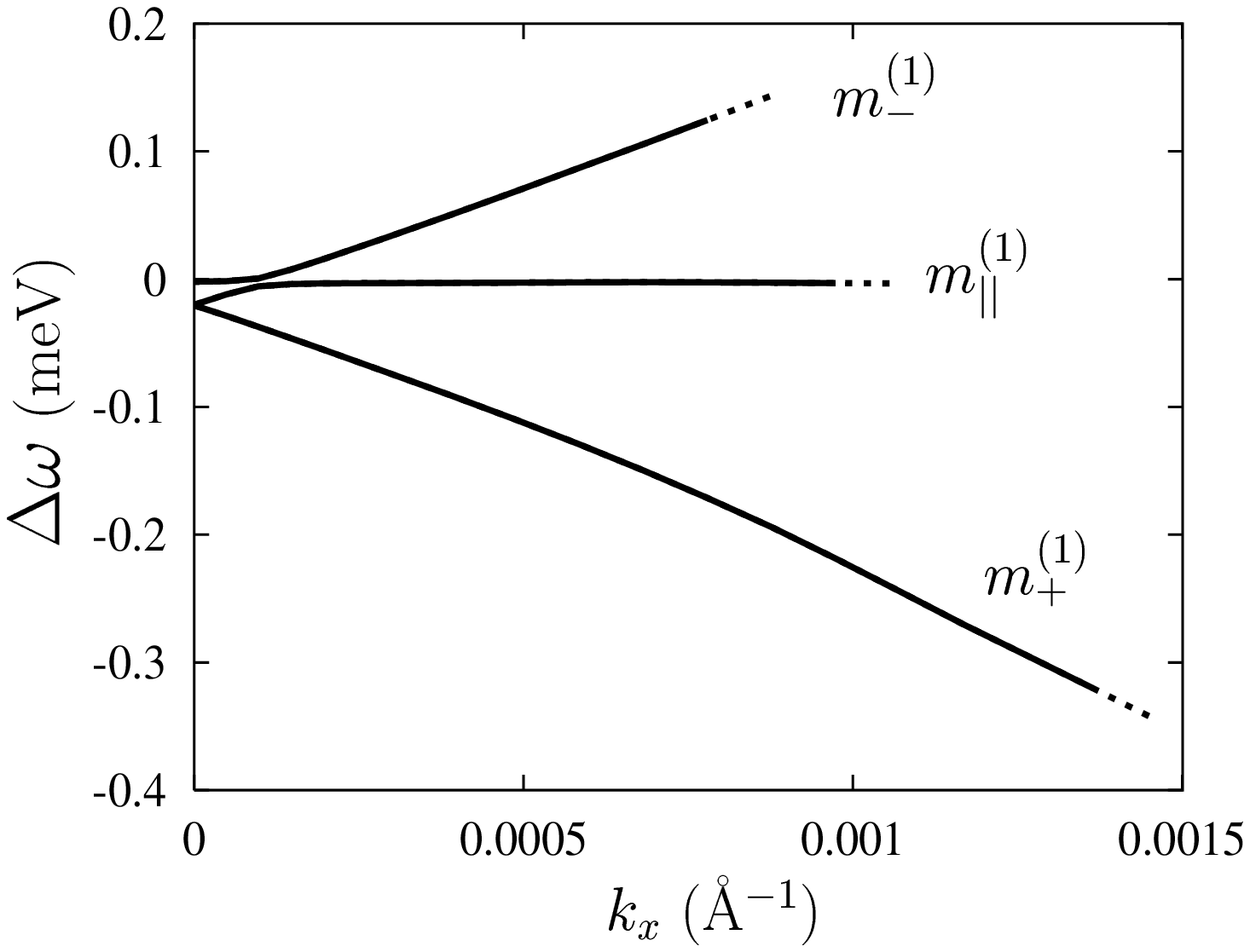}}}
\end{picture}
\caption{\label{figure5} Splitting of the ISB spin plasmon dispersion, for ${\bf k} || [100]$.
$\Delta \omega$ denotes the difference of the spin plasmon frequencies with and without crystal magnetic
fields. The dots indicates that the plasmons enter the particle-hole continuum and become subject to Landau
damping.
}
\end{figure}

Despite the presence of the crystal magnetic field (Fig. \ref{figure3}), the spin plasmons emerge
as distinct collective excitations, i.e.  sharp lines with long lifetimes (limited only by other scattering
mechanisms such as impurities or phonons). Thus, the D'yakonov-Perel' decoherence mechanism does not directly affect
the spin plasmon lifetimes. However, the crystal magnetic fields cause a broadening of the 
particle-hole continuum, as evident from its finite width at $\kp=0$
(a similar effect would be caused by including band-nonparabolicity \cite{warburton}).
This broadening of the particle-hole continuum has an indirect effect on the spin plasmon lifetimes,
in the sense that it may slightly enhance the effectiveness of extrinsic or phonon scattering.
Notice that in our system the plasmon frequencies are comparable to the LO phonon frequency in GaAs
(35.6 meV). In practice, a somewhat wider quantum well may thus be preferable to reduce line broadening
due to phonon scattering.

A much more pronounced signature of the BIA and SIA crystal magnetic fields appears in the 
spin plasmon dispersions themselves, in the form of a splitting into three branches, for finite in-plane wave vector.
This is shown in detail in Fig. \ref{figure5}, where $\Delta\omega$ denotes the difference of the 
spin plasmon frequencies with and without the crystal magnetic fields (in the latter case, the three
spin plasmon branches -- one longitudinal and two transverse -- are identical).
We denote the three branches by $m^{(1)}_+$, $m^{(1)}_{||}$, and $m^{(1)}_-$.
The $m^{(1)}_{||}$ branch has an essentially flat dispersion, after passing through an avoided
crossing for small $k_x$.
The splitting between $m^{(1)}_+$ and $m^{(1)}_-$ is very nearly linear in $\kp$, and reaches values of
about 0.3 meV shortly before the $m^{(1)}_-$ branch enters the particle-hole continuum.

\begin{figure}
\unitlength1cm
\begin{picture}(5.0,7.3)
\put(-6.6,-11.3){\makebox(5.0,7.3){
\includegraphics{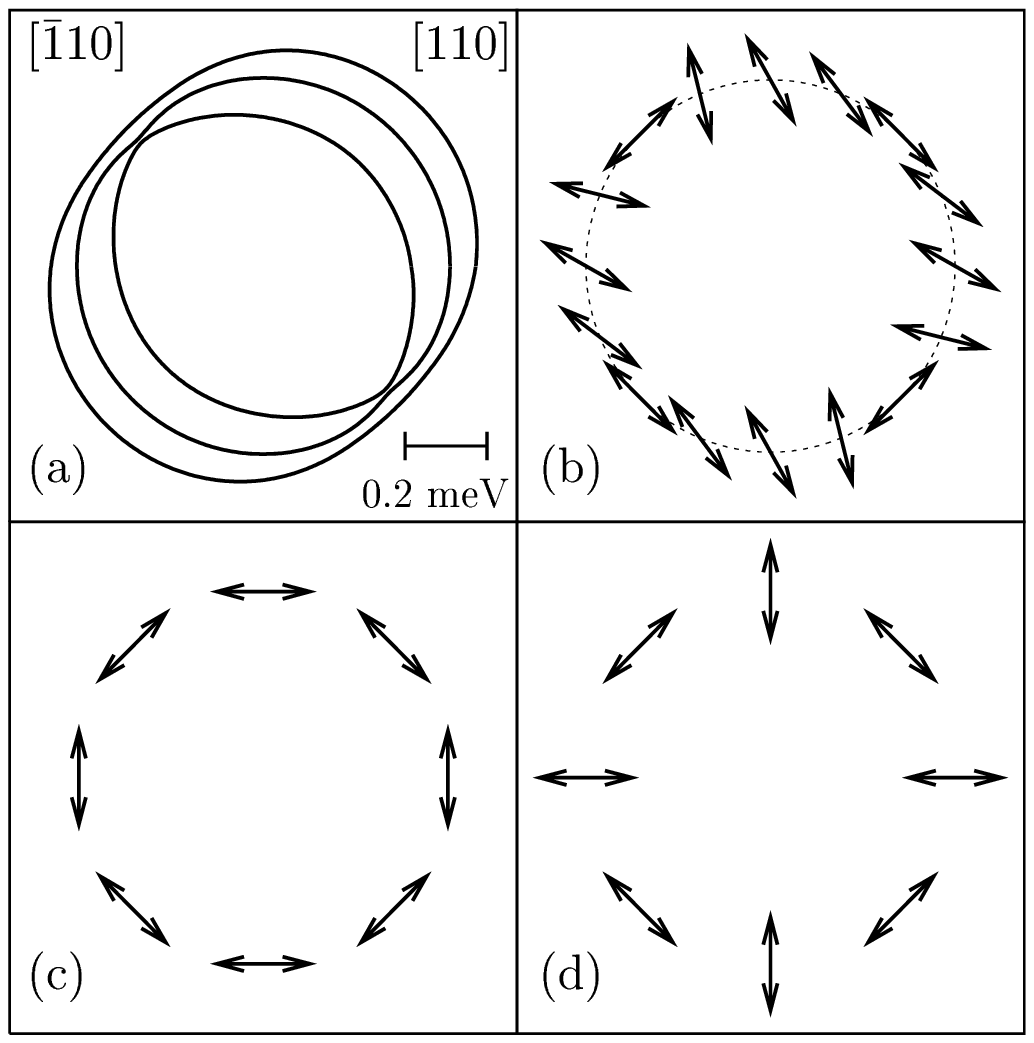}}}
\end{picture}
\caption{\label{figure6} (a) Splitting of the ISB spin plasmon energies as a function of the direction of $\kkp$, for 
$\kp=0.0005 \: \mbox{\AA}^{-1}$, given as the distance between the inner, middle and outer curves, representing the 
$m^{(1)}_+$, $m^{(1)}_{||}$, and $m^{(1)}_-$
branches. The splitting is maximal (0.22 meV) along $[110]$ and $[\bar{1}\bar{1}0]$,
and minimal (0.06 meV) along $[\bar{1}10]$ and $[1\bar{1}0]$. The
polarization of the $m_{||}$ mode for different directions of $\kkp$ is shown in (b) (both SIA and BIA),
(c) (SIA only) and (d) (BIA only).
In the latter cases, the $m_{||}$ mode is strictly in-plane linearly polarized.
}
\end{figure}

The splitting of the spin plasmon dispersions turns out to be highly anisotropic, as shown
in Fig. \ref{figure6}a for in-plane wave vectors of magnitude $\kp=0.0005 \:\mbox{\AA}^{-1}$ and
different directions. The radial distance between the three concentric curves represents the splitting
between $m^{(1)}_+$, $m^{(1)}_{||}$, and $m^{(1)}_-$. The splitting is maximal (0.22 meV) along the $[110]$ 
and  $[\bar{1}\bar{1}0]$ directions, and minimal (0.06 meV) along $[\bar{1}10]$ and $[1\bar{1}0]$.
In the latter cases, we observe an avoided crossing between the 
$m^{(1)}_{||}$ and $m^{(1)}_+$ branches.

The selection rules for inelastic light scattering, see Eq. (\ref{III.8}), are tied to the polarizations of 
the three spin plasmon modes. Figs. \ref{figure6}b-d illustrate the polarization of the $m^{(1)}_{||}$
branch for different directions of $\kkp$, in the case of (b) both SIA and BIA [which corresponds to
Fig. \ref{figure6}a], (c) SIA only, and (d)
BIA only. In the limiting cases (c) and (d), $m^{(1)}_{||}$ is a linearly polarized, purely in-plane mode,
and $m^{(1)}_{+}$ and $m^{(1)}_-$ are out-of-plane, circularly polarized modes (the SIA-only
case was discussed in Ref. \onlinecite{ullrichflatte}). 
In  case (b) where both SIA and BIA are present, the $m^{(1)}_{||}$ branch is no longer purely in-plane
linearly polarized, but features some admixture of circular polarization
(we only plot the in-plane linearly polarized component). This admixture is very small
(less than 1 \%) for most orientations, but becomes substantial in the vicinity of the avoided crossings
along $[\bar{1}10]$ and $[1\bar{1}0]$. 
The other two modes, $m^{(1)}_{+}$ and $m^{(1)}_-$, are again
orthogonal to $m^{(1)}_{||}$, i.e. mostly out-of-plane, circularly polarized.

The splitting at finite $\kkp$ between the $m^{(1)}_{+}$ and $m^{(1)}_-$ modes 
can be explained by a {\em precession} in two opposite directions
of the magnetization orientation 
of the SDE about a uniform effective magnetic field, $\Delta {\bf B}_{\rm eff}(\kkp)$.
In the limiting cases (c) and (d) (SIA and BIA only), the direction of this effective field 
is determined by  the difference of the crystal magnetic fields of the first and second subband,
$\Delta {\bf B}_{\rm eff}(\kkp) \sim {\bf B}_{\rm eff,2}(\qqp + \kkp) - {\bf B}_{\rm eff,1}(\qqp)$.
Case (b) (both SIA and BIA) is more subtle since  ${\bf B}_{\rm eff,2} -
{\bf B}_{\rm eff,1}$ now depends on both $\kkp$ and $\qqp$. Of course, a uniform $\Delta {\bf B}_{\rm eff}(\kkp)$
can still be {\em defined} from the polarization and the splitting of the $m^{(1)}_{+}$ and $m^{(1)}_-$ modes.
Compensating the $\qqp$-dependence of ${\bf B}_{\rm eff,2} - {\bf B}_{\rm eff,1}$ therefore
requires an additional, $\qqp$-dependent contribution to $\Delta {\bf B}_{\rm eff}(\kkp)$, 
which is provided by collective (dynamical xc) effects.

\section{Conclusions}\label{secIV}

In this paper, we have studied ISB spin dynamics in a quantum well in the presence
of bulk and structural inversion asymmetry (BIA and SIA). We have found that there is a
unique signature of inversion asymmetry and spin-orbit coupling: the ISB spin plasmons exhibit a three-fold
splitting. The magnitude of this splitting depends both on the magnitude and orientation of
the in-plane wave vector associated with the spin plasmons. As a result of the interplay
of BIA and SIA, we find a pronounced anisotropy of the ISB spin plasmon splitting.
This anisotropy is of a different nature than the anisotropy that was previously seen
in the spin-flip Raman spectra of {\em intra}subband spin excitations \cite{jusserand}
(see Fig. \ref{figure2}), since it arises from the difference of the crystal magnetic fields
of {\em two} subbands. Thus, the predicted anisotropic ISB spin plasmon spin splitting 
should be a sensitive test for many-body theories of electronic and spin excitations in semiconductor
nanostructures.

Another conclusion of this paper points to the dominant role of collective effects in the 
ISB dynamics in quantum wells. It turns out that the D'yakonov-Perel' decoherence mechanism
is not effective in limiting the lifetime of ISB spin plasmons, which are expected to be
rather long-lived in the absence of other efficient scattering mechanisms. In our previous paper
\cite{ullrichflatte} we had come to the same conclusion for the more specialized case of 
Rashba splitting only. We have, therefore, extended that result in this paper to both BIA and SIA.

In the materials considered here (GaAs/$\rm Al_{0.3}Ga_{0.7}As$), the effects of
spin-orbit coupling are relatively weak compared to other materials such as InAs or AlSb,
which can be expected to produce a much more pronounced spin plasmon splitting.
In those cases, however, the $2\times 2$ conduction band Hamiltonian used in this paper can no longer be 
expected to be accurate enough: for instance, band nonparabolicity will become important.
Thus, a more detailed treatment of band structure will be needed, which will be the subject
of future studies.

\begin{acknowledgments}
This work was supported in part by DARPA/ARO DAAD19-01-1-0490
and by the University of Missouri Research Board.
\end{acknowledgments}

\end{document}